\documentclass[12pt,epsf,amstex]{article}
\usepackage [dvips]{graphicx}
\usepackage{amsmath}
\usepackage{amssymb}
\usepackage{epsfig}
\usepackage{verbatim}
\usepackage{color}

\addtocounter{secnumdepth}{1}
\font\capfont=cmbx12 at 50 pt 
\newbox\capbox \newcount\capl \def\a{A}
\def\docappar{\medbreak\noindent\setbox\capbox\hbox{%
\capfont\a\hskip0.15em}\hangindent=\wd\capbox%
\capl=\ht\capbox\divide\capl by\baselineskip\advance\capl by1%
\hangafter=-\capl%
\hbox{\vbox to8pt{\hbox to0pt{\hss\box\capbox}\vss}}}
\def\cappar{\afterassignment\docappar\noexpand\let\a }

\begin{document}

\newcommand{\ee}{{\rm e}}
\newcommand{\dd}{{\rm d}}
\newcommand{\p}{\partial}

\newcommand{\bA}{\mathbf{A}}
\newcommand{\bB}{\mathbf{B}}
\newcommand{\bc}{\mathbf{c}}
\newcommand{\bC}{\mathbf{C}}
\newcommand{\bF}{\mathbf{F}}
\newcommand{\ff}{\mathbf{f}}
\newcommand{\bM}{\mathbf{M}}
\newcommand{\bO}{\mathbf{O}}
\newcommand{\bq}{\mathbf{q}}
\newcommand{\br}{\mathbf{r}}
\newcommand{\bR}{\mathbf{R}}
\newcommand{\bs}{\mathbf{s}}
\newcommand{\bS}{\mathbf{S}}
\newcommand{\bT}{\mathbf{T}}
\newcommand{\bv}{\mathbf{v}}
\newcommand{\bx}{\mathbf{x}}

\newcommand{\avA}{{\la A \ra}}
\newcommand{\avP}{{\la P \ra}}
\newcommand{\avS}{{\la S \ra}}
\newcommand{\avV}{{\la V \ra}}

\newcommand{\AnE}{{A}_{n_E}}
\newcommand{\LnE}{{L}_{n_E}}
\newcommand{\PnE}{{P}_{n_E}}
\newcommand{\SnF}{{S}_{n_F}}
\newcommand{\VnF}{{V}_{n_F}}
\newcommand{\nF}{{n_F}}
\newcommand{\RnF}{R_{n_F}}
\newcommand{\nE}{{n_E}}
\newcommand{\RnE}{R_{n_E}}
\newcommand{\wnF}{w_{n_F}}
\newcommand{\rh}{r}
\newcommand{\sfrh}{{\sf r}}
\newcommand{\sfrhm}{{\sf r}_{\rm max}}   
\newcommand{\sfRh}{{\sf R}}
\newcommand{\pp}{{\prime\prime}}

\newcommand{\cp}{{\cal P}}
\newcommand{\barp}{\bar{\cp}}
\newcommand{\tf}{\tilde{f}}

\newcommand{\sgn}{\mbox{sgn}}
\newcommand{\cst}{\mbox{cst}}

\newcommand{\hh}{\frac{1}{2}}
\newcommand{\la}{\langle}
\newcommand{\ra}{\rangle}
\newcommand{\beq}{\begin{equation}}
\newcommand{\eeq}{\end{equation}}
\newcommand{\bea}{\begin{eqnarray}}
\newcommand{\eea}{\end{eqnarray}}
\def\lsim{\:\raisebox{-0.5ex}{$\stackrel{\textstyle<}{\sim}$}\:}
\def\gsim{\:\raisebox{-0.5ex}{$\stackrel{\textstyle>}{\sim}$}\:}

\numberwithin{equation}{section}
\thispagestyle{empty}
\title{{\Large  
{\bf Many-faced cells and many-edged faces\\[2mm] 
in 3D Poisson-Voronoi tessellations}\\[2mm]
}
} 

\author{{H.J.~Hilhorst$^1$  and E.A.~Lazar$^2$}\\[5mm]
{\small $^1$ Laboratoire de Physique Th\'eorique, B\^atiment 210}\\
{\small Universit\'e Paris-Sud and CNRS,
91405 Orsay Cedex, France}\\
{\small $^2$ Materials Science and Engineering, University of
  Pennsylvania}\\ 
{\small Philadelphia, PA 19104, USA}}

\maketitle

\begin{small}
\begin{abstract}
\noindent

Motivated by recent new Monte Carlo data we investigate
a heuristic asymptotic theory that applies to
$n$-faced 3D Poisson-Voronoi cells in the limit of large $n$.
We show how this theory may be extended to $n$-edged cell {\it faces}.  
It predicts the leading order large-$n$ behavior
of the average volume and surface area of the $n$-faced cell, and
of the average area and perimeter of the $n$-edged face.
Such a face is shown to be surrounded by a toroidal region of
volume $n/\lambda$ (with $\lambda$ the seed density) that is void of seeds.
Two neighboring cells sharing an $n$-edged face are found to
have their seeds at a typical distance
that scales as $n^{-1/6}$ and whose probability law we determine.
We present a new data set of $4 \times 10^9$ Monte Carlo generated 
3D Poisson-Voronoi cells, larger than any before.  
Full compatibility is found between the Monte Carlo data
and the theory. 
Deviations from the asymptotic predictions 
are explained in terms of subleading corrections whose powers in
$n$ we estimate from the data.\\

\end{abstract}
\end{small}
\vspace{12mm}

\noindent LPT-Orsay-14-34
\newpage


\section{Introduction} 
\label{sec_introduction}

\cappar Spatial tessellations
are of interest because of their wide applicability.
The perhaps simplest model of a disordered cellular structure 
is the Poisson-Voronoi tessellation 
obtained by constructing 
Voronoi cells around point-like `seeds' distributed   
randomly and uniformly in space.
Whereas two- and three-dimensional Poisson-Voronoi cells are relevant for
real-life cellular structures, the higher-dimensional case has
applications in data analyses of various kinds. 
An excellent overview of the many applications is given in the
monograph by Okabe {\it et al.} \cite{Okabeetal00}. 
\vspace{2mm}

Beginning with the early work of Meijering \cite{Meijering53}, 
much theoretical effort has been spent on finding exact analytic
expressions for the basic
statistical properties of the Voronoi tessellation, in particular in
spatial dimensions $d=2$ and $d=3$.
Quantities of primary interest 
are the probability $p_n(d)$ that a cell have exactly $n$
sides (in dimension $d=2$) or $n$ faces (in dimension $d=3$).
Among the very few analytic results that are available for these quantities,
there is a determination \cite{Hilhorst05a,Hilhorst05b} of the 
asymptotic behavior of $p_n(2)$ in the large-$n$ limit.
That calculation also yields the asymptotic behavior of the average
area and perimeter of the two dimensional $n$-sided cell.
Following that exact work 
a heuristic theory was developed \cite{Hilhorst09},
valid again in the large-$n$ limit,
that for $d=2$ reproduces the exact
results and that may also be applied in dimension $d>2$.
In this work we will confront the predictions of this 
`large-$n$ theory', as we will call it, with newly obtained Monte
Carlo data on 3D Poisson-Voronoi cells.

Large-$n$ theory is based on the idea that certain properties
of a large $n$ cell, just like those of a statistical system
in the thermodynamic limit, 
acquire sharply peaked probability distributions
that may for many purposes be replaced with their averages.
We will be interested in
the most characteristic cell properties, {\it viz.}
the average volume $V_\nF$ and surface area $S_\nF$ of an $\nF$-faced cell,
and the average area $A_\nE$ and perimeter $P_\nE$ of an $\nE$-edged 
cell face. 
Large-$n$ theory assumes that for
$\nF\to\infty$ the $\nF$-faced cell tends to a sphere and
predicts the leading asymptotic
behavior of $V_\nF$ and $S_\nF$, {\it viz.}
power laws in $\nF$, including their prefactor.
We here extend this theory such as to
also make predictions for $A_\nE$ and $P_\nE$ as $\nE\to\infty$.

It appears that in the case of the many-edged face 
an important role is played by the distance, to be called $2L$, 
between the seeds of the cells sharing that cell face.
We will refer to $L$ as the `focal distance' because of a
superficial 
resemblance to the foci of, {\it e.g.,} an ellipse.
The extended theory provides an expression for  the probability
distribution of $L$ given $\nE$. 
It appears that whereas $\AnE$  and $\PnE$
increase with $\nE$, the average focal distance $\LnE$ {\it decreases\,} to
zero as $\nE\to\infty$.
\vspace{2mm}

Monte Carlo simulation of Poisson-Voronoi cells
has a tradition that is many decades old.
A computer code developed by Brakke
\cite{Brakke8x} in the 1980's is still used today.
The quality of a Monte Carlo
simulation is first of all determined by the
number of cells that it has generated.

Recent Monte Carlo work by Mason {\it et al.} \cite{Masonetal12}
and by Lazar {\it et al.} \cite{Lazaretal13} focused on the
statistical topology of networks in two and three dimensions.
In Ref.\,\cite{Lazaretal13} Lazar {\it et al.}, using Brakke's code,
produced a data set of 250 million three-dimensional Poisson-Voronoi
cells, larger than any ever obtained before.
The simulation generates successive batches of $10^6$ cells from
$10^6$ seeds randomly and uniformly distributed in a cubic volume
with periodic boundary conditions.
The authors provided an analysis of their data%
\footnote{Available on the Internet \cite{website}.}
with strong emphasis on the identification of the frequency of
different topological cell types.

In the present work we extend the data set to 
four billion ($4 \times 10^9$) three-dimensional cells.
We then compare this enlarged data set to large-$n$ theory.
We find that in all cases the Monte Carlo data are fully compatible
with the predictions of the theory.
There appear to be significant large
finite size corrections.
We discuss to what extent the theoretical law for these subleading terms
may be inferred from the data.
\vspace{2mm}

This paper is organized as follows.
In section \ref{sec_cell}
we consider first the theory and then the Monte Carlo data for the
$\nF$-faced cell.
In section \ref{sec_facetheory}
we extend the theory to the $\nE$-edged cell face and in section 
\ref{sec_faceMC} we present and discuss the Monte Carlo data for
those faces.
In section \ref{sec_higher} we consider subleading terms to the asymptotic
behavior.
In section \ref{sec_discussion} we present a table with our main
results and a critical dicussion of their validity.
In section \ref{sec_conclusion} we conclude.



\section{The many-faced cell}
\label{sec_cell}

\subsection{Theory and simulations}
\label{sec_thsim}

Let there be a three-dimensional Poisson-Voronoi tessellation of seed
density $\lambda$. We will take $\lambda=1$ unless stated otherwise.
Large-$n$ theory as described in Ref.\,\cite{Hilhorst09}
is directly applicable to the volume and surface area of
the three-dimensional $\nF$-faced cell.
We will simply state the results for these quantities and delve deeper
into the theory only in section \ref{sec_facetheory}. 
When $\nF$ gets large, and if we assume that the cell tends towards a 
sphere%
\footnote{This is a very natural idea.
The approach of large 2D cells to circles, and
  higher-dimensional generalizations of this property, have been
  proved rigorously in the mathematical literature 
\cite{CalkaSchreiber05,HugSchneider07},
albeit under hypotheses that do not cover our case.}
of an as yet unknown radius $\RnF$,
the first neighbor seeds must lie close to 
a spherical surface of radius $2\RnF$.
It was shown in Ref.\,\cite{Hilhorst09}
that the volume enclosed by this spherical
surface must be such that under unconstrained conditions
it would have contained on average $\nF$ seeds, that is,
\beq
\frac{4\pi}{3}(2\RnF)^3 \simeq \nF.   
\label{relRncell}
\eeq
Throughout, the sign `$\simeq$' will denote
an equality valid asymptotically
in the limit $\nF\to\infty$.
Eq.\,(\ref{relRncell}) yields $\RnF$ as a function of $\nF$.
The Voronoi cell of the central seed then has a volume $\VnF$
and surface area $\SnF$ given by%
\footnote{We let $X_n=V_n, S_n, A_n, P_n, L_n$ denote averages. When a
distinction is needed we write $X_n^{\rm th}$ for the leading order
theoretical behavior and $X_n^{\rm MC}$ for a Monte Carlo
determination of $X_n$.\label{footnote_one}}
\begin{subequations}\label{resVnSn}
\beq
V_{\nF}^{\rm th} = \frac{4\pi}{3} \RnF^3 \simeq \frac{\nF}{8},
\label{resVn}
\eeq
\beq
S_{\nF}^{\rm th}= 4\pi \RnF^2  \simeq 
\left( \frac{9\pi}{16} \right)^{1/3} \!\nF^{2/3}.
\label{resSn}
\eeq
\end{subequations}
These theoretical averages 
have been obtained without the aid of any adjustable parameter.

In figure \ref{fig_cell} we have presented the Monte Carlo data for
$V_\nF^{\rm MC}$ and $S_\nF^{\rm MC}$ 
obtained by averaging over a set of four billion ($4 \times 10^9$) cells. 
Each quantity has been divided by its theoretical 
large-$\nF$ behavior (\ref{resVnSn}), 
so that for both the data points are expected to tend
to unity as $\nF\to\infty$.  These data appear to fully conform to this 
limit behavior, even if the finite-$n$ corrections are still large.
We will analyze these subleading terms to the asymptotic laws
in section \ref{sec_higher}. 
\vspace{2mm}

It is worth noting that Eq.\,(\ref{resVn}) generalizes 
Lewis' law \cite{Lewis28} 
for the average area $A^{(2)}_n$ 
of a two-dimensional $n$-sided cell.
This law, inspired a long time ago by the study of
epithelial cucumber cells,
hypothesizes that  $A^{(2)}_n=cn$ with a coefficient $c$ estimated 
in the range from 0.20 to 0.25. 
An exact two-dimensional calculation 
\cite{Hilhorst05b} has shown that this law effectively holds for
2D Poisson-Voronoi cells, albeit only asymptotically, as  
\beq
A^{(2)}_n \simeq \frac{n}{4}\,. 
\label{resA2n}
\eeq
The two-dimensional large-$n$ theory reproduces the exact result
(\ref{resA2n}) and this is one reason why we have confidence that the
three-dimensional relations (\ref{resVnSn}) are also exact.  

\subsection{Comments}
\label{sec_sphericity}

We conclude this section by a few comments. 
\vspace{2mm}

{\it 1. Balance of entropic forces.}
Expression (\ref{relRncell}) results \cite{Hilhorst09}
from a balance between
two `forces,' both of purely entropic origin and extensive in $\nF$.
The first one comes from the necessity -- if there is to be an
$\nF$-faced cell -- to have $\nF$ first-neighbor seeds in the 
vicinity of the central seed; the entropy of such a configuration 
{\it increases\,} with the size of the allowable vicinity.
The second one comes from the necessity for all other seeds not to interfere,
and hence to stay
out of an exclusion volume surrounding this vicinity; the entropy of the
other seeds {\it decreases\,} with growing size of the exclusion volume.
\vspace{2mm}

{\it 2. Local and global deviations from sphericity.}
The statement that the `cell surface tends to a sphere'
may be decomposed into 
(i) `the first-neighbor seeds align along a surface,' and
(ii) `this surface tends to a sphere.'
A few words are in place about both.

(i) The {\it local\,} fluctuations of the first-neighbor positions 
perpendicular to their surface of alignment
is characterized by a width $w_{\nF}$. 
The scaling of $w_\nF$ with $\nF$ results from the entropy balance;
in three dimensions $w_n \sim n^{-2/3}$ was found 
\cite{Hilhorst09}.  

(ii) How closely the surface of alignment
approaches a sphere is determined by its
{\it global\,} properties. It was shown in Ref.\,\cite{Hilhorst05b}
that the surface  of the two-dimensional $n$-sided cell
(actually, a  closed curve)
is subject to `elastic' deformations at the scale of the cell itself,
the elasticity being again of entropic origin.
The elastic entropy remains finite as $n\to\infty$ and
does not weigh in the entropy balance that determines 
the two-dimensional $R_{n}^{(2)}$ and $w_{n}^{(2)}$.
However, the elastic modes do
contribute to the deviations of the surface from
sphericity (actually, circularity in 2D).

For finite $n$ there is no sharp distinction between (i) and (ii),
but in 2D they were shown to decouple when $n\to\infty$. 
\vspace{2mm}

{\it 3. Monte Carlo evidence for the approach to sphericity.}
The fluctuations away from sphericity are still 
fairly large for the values of $\nF$ that appear in the simulations. 
Upon assuming a 3D scenario analogous to the one in 2D we conclude that
these fluctuations are due to a combination of the nonvanishing
shell width $w_\nF$ and the elastic deformations. 

\begin{figure}
\begin{center}
\scalebox{.40}
{\includegraphics{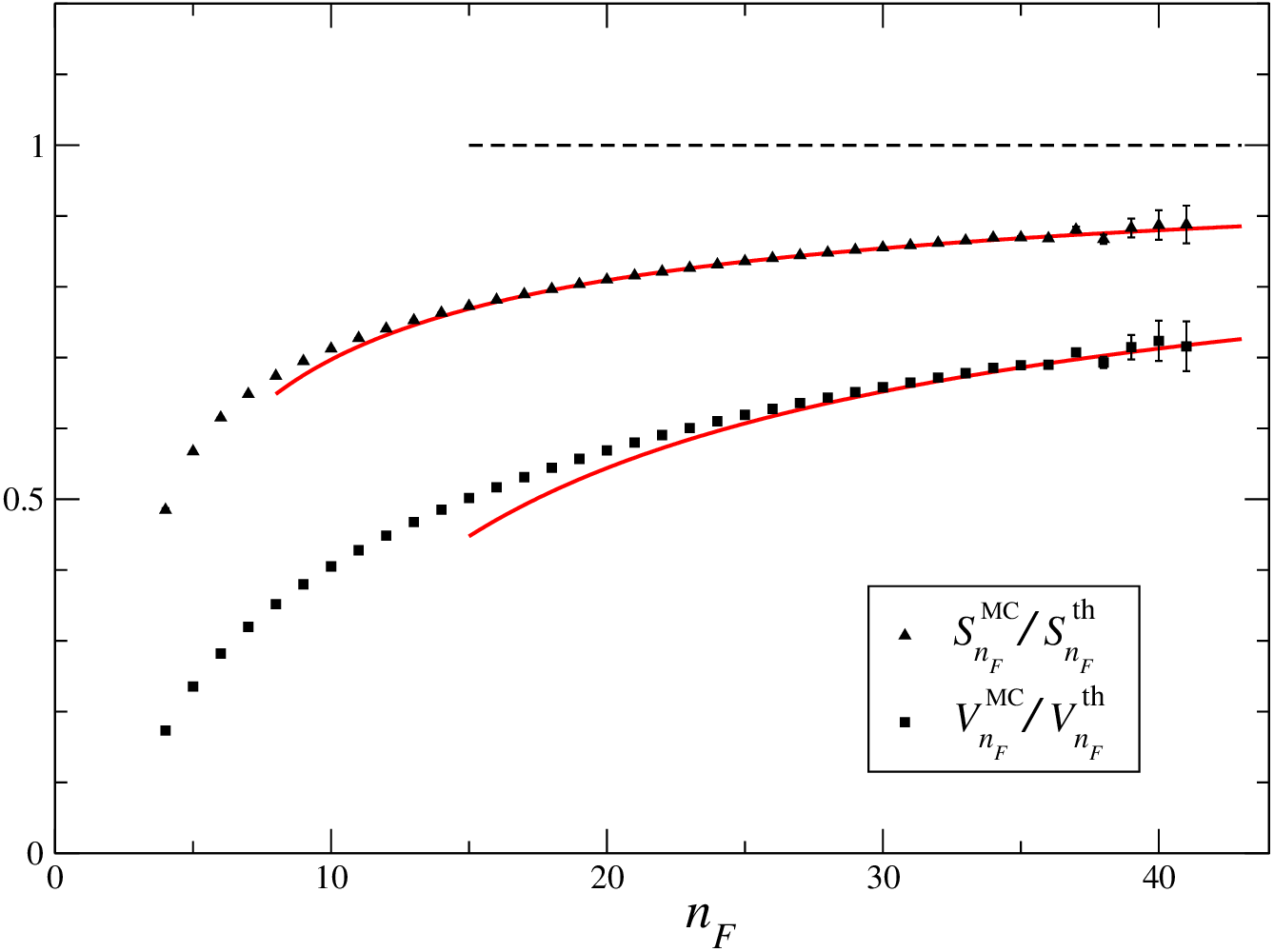}}
\end{center}
\caption{\small Monte Carlo 
averages $V_\nF^{\rm MC}$ and  $S_\nF^{\rm MC}$ of
the volume and surface area, respectively,
of an $\nF$-sided cell, each divided by its theoretical asymptotic
behavior, Eqs.\,(\ref{resVnSn}). Both sets of data points 
are predicted, therefore, to tend to unity as $\nF\to\infty$.
The solid red lines approach this limit value
as  $\sim n^{-2/3}$ and represent our best estimates for the
next-order correction to the leading asymptotic behavior 
(section \ref{sec_higher}).}
\label{fig_cell}
\end{figure}

The Monte Carlo results confirm, however,
the hypothesized approach to sphericity for the following reason.
From Fig.\,\ref{fig_cell} and the known values (\ref{resVnSn}) of 
$V_\nF^{\rm th}$ and $S_\nF^{\rm th}$
one sees that the ratio $6\pi^{1/2}V_\nF^{\rm MC}/(S_\nF^{\rm MC})^{3/2}$ 
tends to unity when $\nF\to\infty$.
If $S_\nF^{\rm MC}$ referred to a single surface enclosing a volume
$V_\nF^{\rm MC}$, this ratio could be unity
only if that surface enclosed the largest possible volume, 
that is, if it were a sphere. 
For the sharply peaked distribution of surface areas
observed in our simulations the same conclusion remains valid.
\vspace{2mm}

{\it 4. Entropy balance and elastic modes.}
The nonextensivity of the elastic entropy allows for the entropy balance
to be set up without taking into account the elastic modes, that is, by
considering the surface of alignment as a sphere right from the start.
In the same spirit, when in the next section we will 
consider seed positions that align along
a toroidal surface, we will do so without
regard for the elastic deformations of that surface.

\section{The many-edged face:  theory}
\label{sec_facetheory}

\subsection{Torus}
\label{sec_facetorus}

\subsubsection{Preliminaries}
\label{sec_torus}

\begin{figure}
\begin{center}
\scalebox{.40}
{\includegraphics{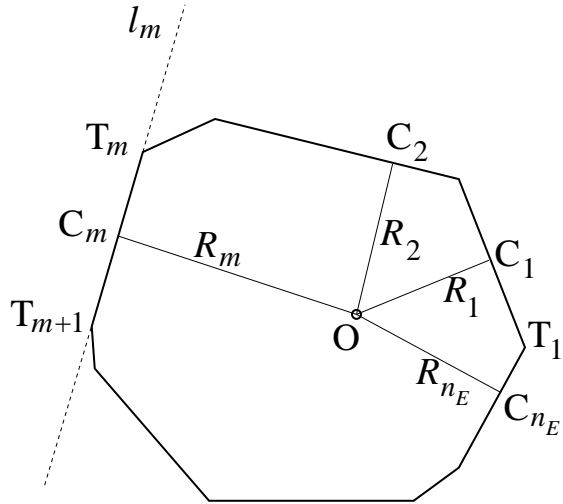}}
\end{center}
\caption{\small Geometry in the plane (`$xy$' plane) 
  of the $\nE$-edged face shared by two cells having their
  seeds in $\bS_1$ and $\bS_2$. The
  line segment connecting these seeds is perpendicular to this plane
  and is bisected by it in $\bO$. 
  The $m$th edge of the face connects the vertices $\bT_m$ and
  $\bT_{m+1}$  and lies on a line $\ell_m$.
  The $\bC_m$ are the projections of $\bO$ onto the $\ell_m$.}
\label{fig_OCC}
\end{figure}

Let us consider an arbitrarily selected $\nE$-edged cell face
between two neighboring Voronoi cells.
Let the seeds
of the two cells (the `focal' seeds) have positions $\bS_1$ and $\bS_2$. 
By a suitable choice of the origin $\bO$ and the direction
of the $z$ axis we obtain $\bS_1=(0,0,L)$ and $\bS_2=(0,0,-L)$,
where $L$ is the `focal distance'. 
It is a random variable whose distribution we do not know {\it a priori}. 
The $\nE$-edged face is then located in the $xy$ plane; 
a typical face is shown schematically in figure \ref{fig_OCC}.
We number its edges by $m=1,2,\ldots,\nE$ according to increasing
polar angle and let $\ell_m$ denote the line that prolongs the $m$th edge.
We let furthermore 
$\bC_m$ denote the projection of the origin $\bO$ onto $\ell_m$
and $\bT_1,\ldots,\bT_n$ the vertices of the $\nE$-edged face.

\begin{figure}
\begin{center}
\scalebox{.45}
{\includegraphics{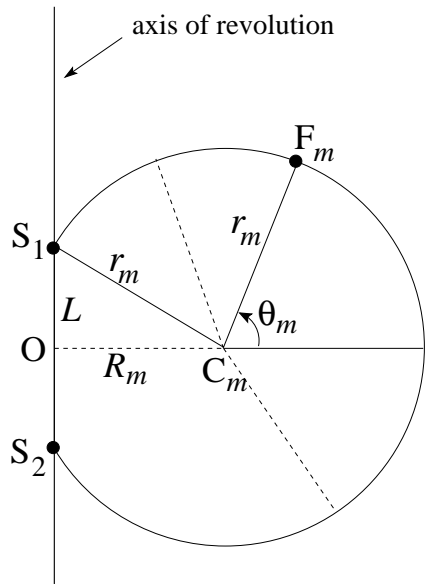}}
\end{center}
\caption{\small Geometry in the first-neighbor plane
  plane passing through the seeds $\bS_1$, $\bS_2$, and
  $\bF_m$. Point $\bC_m$ is the center of the circle passing through
  these three seeds. The cell face studied lies in the plane through
  O perpendicular to the axis of revolution (the `$z$' axis).
  Rotating the circular arc shown about this axis produces a spindle
  torus: its minor radius $r_m$ is larger than its major radius $R_m$.
  Each dashed line lies in a plane equidistant to two of the three seeds.}
\label{fig_SSF}
\end{figure}

The $m$th edge is common to the Voronoi cells of $\bS_1$, $\bS_2$, and
of a third seed whose position we call $\bF_m$. 
We will refer to the $\bF_m$ as the `first neighbors' of the
pair $(\bS_1,\bS_2)$.
Figure \ref{fig_SSF} represents the plane through these three seeds,
that we will also refer to as the $m$th `first-neighbor' plane. 
The three planes that perpendicularly bisect the line segments
connecting these three seeds intersect along line $\ell_m$.
This line is perpendicular to the  $m$th first-neighbor plane 
and intersects it in
$\bC_m$, which is therefore equidistant to the three seeds, as shown by
the large circular arc of radius $r_m$. 
As announced at the end of section \ref{sec_cell}, we are assuming
that it is safe in this discussion to neglect
the elastic deformations of the torus.

\subsubsection{Large-$n$ limit}
\label{sec_toruslargen}

For the cell face of figures \ref{fig_OCC} and \ref{fig_SSF}
we now develop the following extension of the
large-$n$ theory. To simplify notation we write $n$ instead of $\nF$. 
Let us consider the subset of faces with fixed focal distance $L$.
It is natural to assume that in the limit of large $n$ the area of the
$n$-edged face will grow without limit and that its shape
will approach a circle
of some as yet unknown radius that we will call $\sfRh_n$.
More precisely, all  $R_m/\sfRh_n$ will tend to unity%
\footnote{Almost surely, in the mathematical sense.}
when $n\to\infty$.
According to figure \ref{fig_SSF}
there must then also be an $\sfrh_n$ related to $\sfRh_n$ by
\beq
\sfrh_n^2=\sfRh_n^2+L^2
\label{relRrL}
\eeq 
and which is such that $r_m/\sfrh_n$ will tend to unity when $n\to\infty$.
In that limit, as $m$ varies from $1$ to $n$, 
the large circular arc in figure \ref{fig_SSF} turns
around the axis of revolution and describes a torus 
whose major and minor radii are $\sfRh_n$ and $\sfrh_n$. 
Since $\sfRh_n \leq \sfrh_n$,
this torus has no hole and is actually a spindle torus. 
The $\bF_m$ lie close to the surface of this torus%
\footnote{The surface of a spindle torus is called an `apple'.}
in a thin shell whose width $w_n$ vanishes with growing $n$.
There can be no seeds inside this torus
as this would destroy the $n$-edgedness of the face.


\subsection{Probability $\cp_n$ of occurrence of an $n$-edged face}
\label{sec_pn}

Given two adjacent cells that share an $n$-edged face,
we now ask for the probability $\cp_n$
that the two focal seeds be at distance $2L$
{\it and\,} that the $n$ first neighbor seeds 
be located in a toroidal shell with minor radius $\sfrh$,
and therefore with major radius $\sfRh=(\sfrh^2-L^2)^{1/2}$.
It will have advantages to express $\cp_n$ as a function
of the independent variables $\sfrh$ and
\beq
x = \frac{L}{\sfrh}\,.
\label{defx}
\eeq
Since it is proportional to the number of
microscopic seed configurations compatible with the constraints 
$(n,\sfrh,x)$, and because of the analogy with thermodynamics, we will refer
to $\,\log\cp_n(\sfrh,x)\,$ as an `entropy'. 
We will now determine an explicit although approximate expression
for this entropy and study its variation with $\sfrh$ and $x$.

Let us write $V_0$ for the volume of the torus with parameters $\sfrh$ and
$L$, $S_0$ for its surface area, and 
\beq
V_1=w_nS_0
\label{relV1wnS0}
\eeq
for the volume of the shell of width $w_n$ at the surface of the torus. 
Let $\lambda$ (which may be scaled away) be the three-dimensional
seed density. We then have
\beq
\cp_n(\sfrh,x) \simeq \cst \times (x\sfrh)^2\,
\frac{\ee^{-\lambda V_1}(\lambda V_1)^n}{n!}\, \ee^{-\lambda V_0},
\label{xp}
\eeq
in which, here and henceforth, `cst' stands for a constant that may each
time be a different one, and where
$(x\sfrh)^2=L^2$ is the phase space factor associated with two seeds being at
distance $2L$, the Poisson distribution
$\ee^{-\lambda V_1}(\lambda V_1)^n/n!$
is the probability that in a random seed distribution of density
$\lambda$ the volume $V_1$ contain exactly $n$ seeds,
and $\ee^{-\lambda V_0}$ is the probability that the volume $V_0$ 
contain no seeds. 
Equation (\ref{xp}) is obviously an approximation: 
for one thing, it does not take
into account the detailed individual positions of the first neighbor
seeds in $V_1$, but only restricts them to the shell.
We will take (\ref{xp}) seriously, nevertheless, and see where it leads us.

The expressions, needed in (\ref{xp}), 
for the volume $V_0$ and the surface $S_0$
of the torus with parameters $\sfrh$ and $x=L/\sfrh$ are 
\begin{subequations}\label{xVStorus}
\beq
V_0 = 2 \pi^2 \sfrh^3 g(x),
\label{xVtorus}
\eeq
\beq
S_0 = 4 \pi^2 \sfrh^2 f(x),
\label{xStorus}
\eeq
\end{subequations}
in which
\begin{subequations}\label{xfg}
\beq
\pi f(x) =  x + (\pi-\arcsin x)\sqrt{1-x^2}, 
\label{xf}
\eeq
\beq
\pi g(x) = \pi f(x) - \tfrac{1}{3}x^3.
\label{xg}
\eeq
\end{subequations}
For later use we note the small-$x$ expansions
\bea\label{fgsmallx}
f(x) &=& 1 - \tfrac{1}{2}x^2 + \frac{1}{3\pi}x^3 +{\cal O}(x^4),
\nonumber\\[2mm]
g(x) &=& 1 - \tfrac{1}{2}x^2 + {\cal O}(x^4).
\eea
The shell width $w_n$, also needed in (\ref{relV1wnS0}), is a function
of $\sfrh$ and $x$ that we will determine in the next section.


\subsection{Shell width $w_n$}
\label{sec_wn}

\begin{figure}
\begin{center}
\scalebox{.70}
{\includegraphics{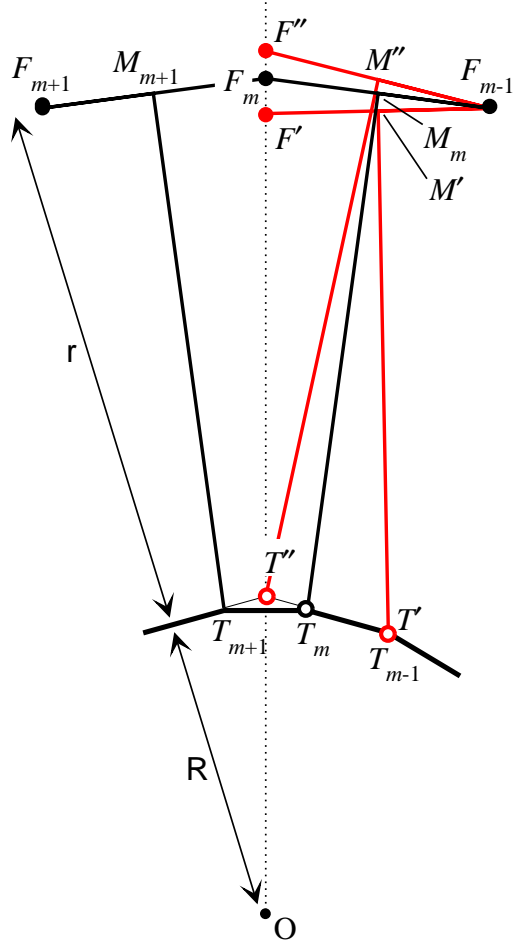}}
\end{center}
\caption{\small Geometry in the plane of the face after
all first neighbor seeds $\bF_j$ have been rotated as explained in
the text. 
The heavy line linking $\ldots,\bT_{m-1},\bT_m,\ldots$ is the face boundary
when the $m$th neighbor is located at $\bF_m$.
When $\bF_m$ moves to $\bF^\prime$ (or to $\bF^\pp$),
then $\bT_m$ moves to $\bT^\prime$ (or to $\bT^\pp$).}
\label{fig_w}
\end{figure}

Our determination of $w_n$ will exploit an invariance 
hidden in this problem.
The $m$th edge of the face is a segment of a line $\ell_m$ that 
is perpendicular to the plane of figure \ref{fig_SSF} and intersects
this plane in $\bC_m$. Along $\ell_m$ the three Voronoi
cells of $\bS_1$, $\bS_2$, and $\bF_m$, join.
The faces separating these cells are located in planes that are also
perpendicular to the plane of figure \ref{fig_SSF} and intersect it along
the dashed lines passing through $\bC_m$.
Suppose now that seed $\bF_m$ moves along the circular arc in figure
\ref{fig_SSF}. This will leave the position of $\bC_m$ invariant;
hence it will leave line $\ell_m$ invariant; and since the set of
lines $\{\ell_m\}$ determines the perimeter of the face, it will leave
the face invariant.
We may therefore rotate all first neighbors $\bF_m$ to a position with
$\theta_m=0$, that is, a position in the plane of the face, without
changing the face. 
Having performed this rotation (without introducing a new symbol for
the rotated $\bF_m$) we obtain the situation of figure \ref{fig_w}.
We are now ready to discuss the width $w_n$.

The filled black dots in figure \ref{fig_w}
are the positions after rotation of the first neighbors
$\bF_m$. For convenience we have chosen them as the
vertices of a regular $n$-gon, supposing that this does not affect the 
argument below in any essential way. The edges of the $n$-gon have
midpoints $\bM_m$. The $\bT_m$ are the vertices of the $n$-edged face
of interest, which is also a regular $n$-gon. The 
$M_mT_m$ are the perpendicular bisectors of the
$F_mF_{m-1}$, where we write here $AB$ for the line segment
connecting the two points $\bA$ and $\bB$.
Suppose now that $\bF_m$ moves along the line
through $\bF'$ and $\bF^\pp$ (both points marked by filled red dots).
The midpoint $\bM_m$ then moves along a parallel line with
corresponding points $\bM'$ and $\bM^\pp$. On the left the midpoint
$\bM_{m+1}$ executes the mirrored motion (not shown). As a consequence
line segment 
$T_mT_{m+1}$ is displaced parallel to itself.
When it moves down so far that it passes
through $\bT'$, its neighboring segments disappear; and when it moves
up so high that it passes through $\bT^\pp$, it disappears itself. In both
cases the face ceases to be $n$-edged. The limit points $\bT'$ and
$\bT^\pp$ determine $\bF'$ and $\bF^\pp$. We will identify 
somewhat arbitrarily
the shell width $w_n$ with the segment length
$|F'F^\pp|$, which we calculate as follows.
The angle between $F'F_{m-1}$ and 
$F^\pp F_{m-1}$ is identical to the one between $T'M'$ and 
$T^\pp M^\pp$. All these angles
become very small as $n$ gets large. Neglecting higher order terms in
the angles we have
\beq
\frac{|F'F^\pp|}{|F_mF_{m-1}|} = \frac{|T'T^\pp|}{|T_mM_m|}\,.
\label{ratio}
\eeq
Upon using that the $\bF_m$ and $\bT_m$ are vertices 
of regular polygons and substituting 
$|F_mF_{m-1}|=2\pi(\sfRh+\sfrh)/n$,
$|T'T^\pp|=3\pi\sfRh/n$, and $|T_mM_m|=\sfrh$ we obtain
\bea
w_n(\sfrh,L) &=& \frac{6\pi^2\sfRh(\sfRh+\sfrh)}{n^2\sfrh}  \nonumber\\[2mm]
             &=& \frac{C}{2}(1-x^2+\sqrt{1-x^2})\frac{\sfrh}{n^2}\,, 
\label{xwn}
\eea
in which $C=12\pi^2$ is a constant that will play no role in what follows.
We will write 
\beq
\tf(x) = \frac{1}{2}(1-x^2+\sqrt{1-x^2})f(x),
\label{deftf}
\eeq
so that from relations (\ref{deftf}), (\ref{relV1wnS0}), 
and (\ref{xStorus}) we have
\beq
V_1 = \frac{4\pi^2 C \sfrh^3}{n^2}\tf(x).
\label{xV1}
\eeq
Equations (\ref{xVtorus}) and (\ref{xV1}) 
are the desired expressions for $V_0$ and $V_1$.

\subsection{Analysis of $\cp_n(\sfrh,x)$} 
\label{sec_maximization}

Directly from Eq.\,(\ref{xp}) we have
\beq
\log\cp_n(\sfrh,x) \simeq -\lambda V_1 + n\log\lambda V_1 -\log n! 
- \lambda V_0 + 2\log x - \frac{2}{3} \log\lambda\sfrh^3,
\label{xlogp}
\eeq
which we will study as a function of its two variables.
We may simplify this expression
by noting that in the large-$n$ limit $\lambda V_1$ is
negligible with respect to $\lambda V_0$ and  
$\log\lambda\sfrh^3$ with respect to $n\log\lambda V_1$.
Some further rewriting is useful.
First, we substitute in (\ref{xlogp}) the explicit expressions
(\ref{xVtorus}) and (\ref{xV1}) for $V_0$ and $V_1$.
Second, we may discard from (\ref{xlogp}) 
any terms that do not depend on $\sfrh$ or $x$
and that we may recover later by normalizing the distribution.
Then, instead of $\log\cp_n$ of Eq.\,(\ref{xlogp}),
we may study $\log\barp_n$ given by
\beq
\log\barp_n(\sfrh,x) \simeq n \log\Big( 2\pi^2\lambda\sfrh^3\tf(x) \Big)
- 2\pi^2\lambda\sfrh^3g(x) + 2\log x.
\label{xlogbarp}
\eeq
The first two terms represent two opposing entropic forces
similar to those referred to in section \ref{sec_sphericity} for the case of
the $\nF$-sided cell. 
We are first of all interested in the variation of $\log\barp_n$ with
$\sfrh$.
For fixed $x$, let (\ref{xlogbarp}) be maximal for
$\sfrh=\sfrhm(x)$.
Setting $\partial\log\barp_n/\partial(2\pi^2\lambda\sfrh^3)=0$ 
we obtain 
\beq
2\pi^2\lambda\sfrhm^3(x)g(x) = n.
\label{solrhox1}
\eeq
We now note that in view of (\ref{xVtorus}) the first member of the
above equation is equal to $\lambda V_0$.
Eq.\,(\ref{solrhox1}) therefore says 
that the entropy 
is maximized when the volume of the
torus is such that under unconstrained conditions it would have
contained $n$ seeds. This is the torus counterpart of
Eq.\,(\ref{relRncell}). 

For $n\to\infty$ the maximum in $\sfrh$ corresponds to a narrow peak,
as may be shown by 
an expansion of (\ref{xlogbarp}) about its maximum. The marginal distribution
of $x$, defined as the integral of $\barp_n(\sfrh,x)$ with respect to
its first argument, is therefore obtained by simply taking 
$\sfrh=\sfrh_{\rm max}(x)$ in (\ref{xlogbarp}), which leads to
\beq
\barp_n(\sfrhm(x),x) \simeq \cst \times x^2 
\left( \frac{\tf(x)}{g(x)} \right)^{\! n}.
\label{xbarpm}
\eeq
The ratio $\tf(x)/g(x)$ has its maximum at $x=0$.
Upon expanding for small $x$ with the aid of 
(\ref{xfg}) and (\ref{deftf}) we obtain
\beq
\frac{\tf(x)}{g(x)} = 1 - \frac{3}{4} x^2 + \frac{1}{3\pi} x^3 
+ {\cal O}(x^4).
\label{expfdg}
\eeq
The term of order $x^2$ with the negative coefficient $-3/4$ is
the only one that leaves a trace in the limit $n\to\infty$;
it stems directly from the factor $(1-x^2+\sqrt{1-x^2})/2$ in
(\ref{deftf}), which in turn comes from the shell width. 
Using (\ref{expfdg}) in (\ref{xbarpm}) and letting
$n\to\infty$ we have to leading order
$\barp_n(\sfrhm(x),x) \to \cst\times x^2\exp(-(3n/4)x^2)$, 
so that $x$ is not sharply peaked but has a well-defined distribution
on scale $n^{-1/2}$. 
More precisely, in that limit the scaled variable 
\beq
y= (3\pi n)^{1/2}x/4
\label{defy}
\eeq 
has the distribution $Q(y)$ given by 
\beq
Q(y) = 32\pi^{-2} y^2\exp\left( -\frac{4}{\pi}y^2 \right), \qquad y>0,
\label{xQy}
\eeq
where we have restored the normalization, and where $y$ is such that
its first moment is unity.
\vspace{3mm}

Knowing that $x$ is random on the scale $n^{-1/2}$ we have
from (\ref{fgsmallx}) that $g(x)= 1 + O(n^{-1})$ and 
subsequently from (\ref{solrhox1}) the small-$x$ expansion 
\beq
\sfrhm(x) = \sfrh_n \left[ 1+{\cal O}(n^{-1})\right]
\label{solrmax}
\eeq
with leading order term 
\beq
\sfrh_n = \left( \frac{n}{2\pi^2} \right)^{1/3},
\label{xrn}
\eeq
in which we have set $\lambda=1$. 
In relation (\ref{defx}) we now replace
$\sfrh$ by its leading order value $\sfrh_n$ and obtain, 
also using (\ref{xrn}), 
\bea
L &\simeq& (n/2\pi^2)^{1/3}x \nonumber\\[2mm] 
&=& 2^{5/3}3^{-1/2}\pi^{-7/6}n^{-1/6}y.   
\label{relLy}
\eea
This shows that $L$ varies on scale $n^{-1/6}$.
Since $y$ has unit average we now have for the average $L_n$ of $L$
the expression%
\footnote{See footnote \ref{footnote_one}.}
\beq
L_n^{\rm th} \simeq  2^{5/3}3^{-1/2}\pi^{-7/6}n^{-1/6}.
\label{resLn}
\eeq
Furthermore, as $n\to\infty$ the probability distribution $Q_n$
of the scaled variable $y=L/L_n^{\rm th}$   
is predicted 
to tend to the fixed law $Q(y)$ of Eq.\,(\ref{xQy}).
One may loosely rephrase this scaling with $n^{-1/6}$ by saying that
the many-edgedness of a cell face leads to an attractive force (of
entropic origin) between the two focal seeds.
It was not {\it a priori\,} clear to us that such a phenomenon would occur.
\vspace{3mm}

Knowing now that $L$ is distributed on scale $n^{-1/6}$,
relation (\ref{relRrL}) tells us that 
$\sfrh_n$ and $\sfRh_n$ must be equal to leading order, and hence
\beq
\sfRh_n \simeq \left( \frac{n}{2\pi^2} \right)^{1/3}.
\label{xRn}
\eeq
For the shell width $w_n$ and the shell volume $V_1$
we find with the aid of  (\ref{xRn}), (\ref{xwn}), (\ref{relV1wnS0}), 
and (\ref{xStorus}) the scaling behavior
\beq
w_n \simeq \cst\times n^{-5/3}, \qquad V_1 \simeq\cst\times n^{-1},
\label{xwnaspt}
\eeq
where  we have preferred to
denote the prefactors by `cst'
in view of the arbitrariness in the definition of $w_n$.
Eq.\,(\ref{xwnaspt}) tells us that 
the shell becomes rapidly thinner as $n$ gets larger.
\vspace{3mm}

We finally return to the averages $A_n$ and $P_n$.
Having determined that for $\nE\to\infty$
the $\nE$-edged cell face tends to a circle of a now known radius $\sfRh_n$
we conclude that%
\footnote{See footnote \ref{footnote_one}.}
\begin{subequations}\label{resAnPn}
\beq
A_\nE^{\rm th} = \pi\sfRh_\nE^2 \simeq (4\pi)^{-1/3}\nE^{2/3},
\label{resAn}
\eeq  
\beq
P_\nE^{\rm th} = 2\pi\sfRh_\nE \simeq (4\pi)^{ 1/3}\nE^{1/3}.
\label{resPn}
\eeq
\end{subequations}
These relations are analogous to the laws (\ref{resVnSn}) 
for the cell volume and surface area.
This completes the extension of large-$n$ theory to the $\nE$-edged
cell face in the limit of asymptotically large $\nE$.


\section{The many-edged face: Monte Carlo}
\label{sec_faceMC}

The $4\times 10^9$ cells generated by Monte Carlo simulation
yielded  $N_n$ cell faces of edgedness $n$, adding up to a total 
of $N=\sum_nN_n= 31\,071\,027\,941$ cell faces.
The distribution $N_n$ has been presented in table \ref{table1}
together with our estimates of the fractions $f_n$ of
$n$-edged faces.
In Ref.\,\cite{Lazaretal13} several comparisons with theoretically
known data have been presented as a demonstration that the algorithm
works correctly.
Here we limit ourselves to two such tests, shown in table
\ref{table1b}.
Let $\langle\nF\rangle$ and $\langle\nE\rangle$
stand for the average facedness of a cell and the average edgedness 
of a cell face, respectively. 
The rms deviation of $\nF$ is equal to 3.318,
which leads to an estimate of the standard deviation in its Monte Carlo average
equal to $3.318/\sqrt{4\times 10^9}=0.000\,06$.
The rms deviation of $\nE$ is equal to 1.579,
which leads to an estimate of the standard deviation in its Monte Carlo average
equal to $1.579/\sqrt{N}=0.000\,009$.
The average values from the Monte Carlo simulations together with
these standard deviations
are shown in the first two lines of table \ref{table1b}.
The theoretical values of both averages are exactly known 
(see {\it e.g.} Ref.\,\cite{Okabeetal00}) and shown in the third line.
The agreement between
the Monte Carlo values and these exact results is excellent.

\begin{table}
\begin{center}
\begin{small}
\begin{tabular}{||rrr||rrr||}
\hline
\multicolumn{1}{||r}{$n$}&
\multicolumn{1}{c}{$N_n$}&
\multicolumn{1}{c||}{$f_n$}&
\multicolumn{1}{c}{$n$}&
\multicolumn{1}{c}{$N_n$}&
\multicolumn{1}{c||}{$f_n$}\\
\hline
3   & 4\,187\,261\,126 & $0.134\,764\pm 0.000\,002$ & 12 & 11\,834\,735 
&$(3.809\pm 0.002)\times 10^{-4}\,\,$\\
4   & 7\,140\,019\,564 & $0.229\,797\pm 0.000\,003$ & 13 &  2\,174\,618 
&$(6.999\pm 0.005)\times 10^{-5}\,\,$\\
5   & 7\,505\,993\,048 & $0.241\,575\pm 0.000\,003$ & 14 &     342\,988
&$(1.104\pm 0.002)\times 10^{-5}\,\,$\\   
6   & 5\,914\,222\,488 & $0.190\,345\pm 0.000\,003$ & 15 &      46\,869 
&$(1.508\pm 0.007)\times 10^{-6}\,\,$\\
7   & 3\,621\,030\,915 & $0.116\,540\pm 0.000\,002$ & 16 &       5\,690 
&$( 1.83\pm 0.03)\times 10^{-7}\,\,$\\
8   & 1\,747\,654\,056 & $0.056\,247\pm 0.000\,002$ & 17 &          613 
&$( 1.97\pm 0.08)\times 10^{-8}\,\,$\\
9   &    674\,407\,674 & $0.021\,705\pm 0.000\,001$ & 18 &           41 
&$(  1.3\pm 0.2)\times 10^{-9}\,\,$\\
10  &    211\,374\,682 & $0.006\,803\pm 0.000\,001$ & 19 &            7 
&$(  2.3\pm 0.9)\times 10^{-10}$\\
11  &     54\,658\,826 & $0.001\,759\pm 0.000\,001$ & 20 &            1 
&$(  3\pm3)\times 10^{-11}$\\
\hline
\end{tabular}
\end{small}
\end{center}
\caption{Observed numbers $N_n$ of $n$-edged cell faces 
in a set of $4\times 10^9$ Monte Carlo generated 3D Poisson-Voronoi cells,
and their estimated fractions $f_n$.}
\label{table1}
\end{table}

\begin{table}
\begin{center}
\begin{small}
\begin{tabular}{|l|r@{.}l|r@{.}l|}
\hline
\multicolumn{1}{|l|}{}&
\multicolumn{2}{|c|}{ Expected number $\la\nF\ra$ }&
\multicolumn{2}{|c|}{ Expected number $\la\nE\ra$ }\\
\multicolumn{1}{|l|}{}&
\multicolumn{2}{|c|}{ of faces of a cell }&
\multicolumn{2}{|c|}{ of edges of a face }\\
\hline
Monte Carlo & \phantom{XXXX}15&535\,51   & \phantom{XXXX}5&227\,576    \\
Standard deviation &  0&000\,06   & 0&000\,009    \\
Theory             & 15&535\,457  & 5&227\,573\,4 \\
\hline
\end{tabular}
\end{small}
\end{center}
\caption{Two tests of the Monte Carlo algorithm.}
\label{table1b}
\end{table}

\subsection{Examples of many-edged faces}
\label{sec_examplesMC}

\begin{figure}
\begin{center}
\scalebox{.45}
{\includegraphics{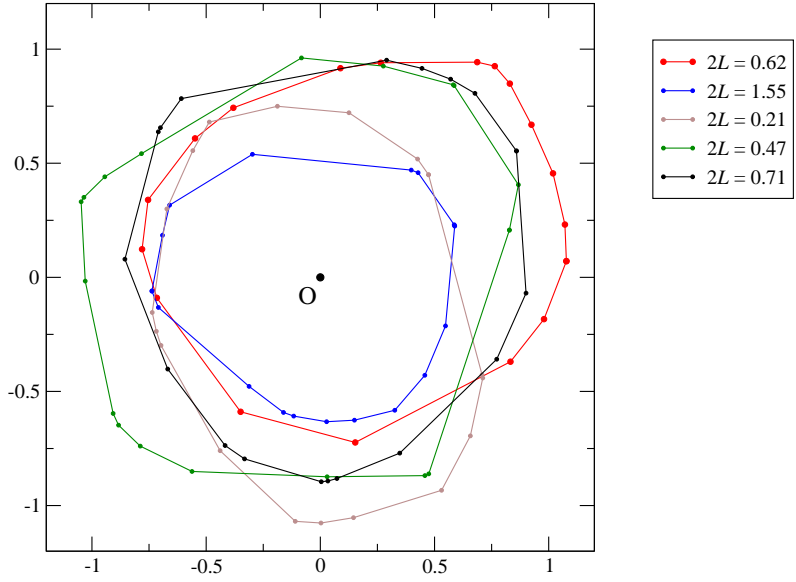}}
\end{center}
\caption{\small Five $18$-sided cell faces found 
in the Monte Carlo simulations of Ref.\,\cite{Lazaretal13},
superposed such that their origins coincide in a single point $\bO$. 
For each, the value $2L$ of the distance between the two focal seeds
is indicated.}
\label{fig_fivefaces}
\end{figure}

In the original
Monte Carlo simulations by Lazar {\it et al.} \cite{Lazaretal13},
that comprised $0.25\times 10^9$ cells,
faces were found with edge numbers up to $\nE=18$.
In figure \ref{fig_fivefaces}
we show the five $18$-edged faces that occurred, superposed
such that their origins coincide.
Some faces, such as the red one, are close to circular,
but the set shows that 
there is still considerable variability in shape and size;
also, the origin, which for $\nE\to\infty$ should be at the center of
the circle, is still fairly eccentric.
It is relevant to recall here that
these same observations held for the many-sided two-dimensional cells
studied in Ref.\,\cite{Hilhorst07}, for which nevertheless an
efficient simulation algorithm has demonstrated the convergence to a circle
at higher values of $n$. 
If the blue face and the gray face
seem to have fewer than $18$ edges, this
is due to some of their vertices 
coinciding at the scale of the figure.

\begin{figure}
\begin{center}
\scalebox{.45}
{\includegraphics{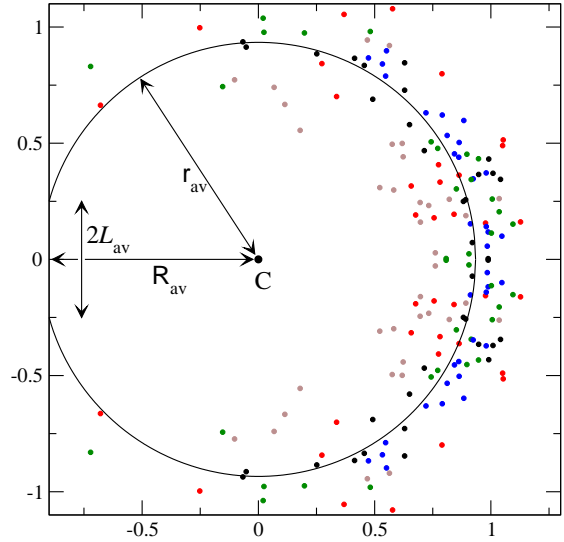}}
\end{center}
\caption{\small Figure based on the same five $18$-sided cell faces as
shown in figure \ref{fig_fivefaces}, with the same color code.
All $5\times 18$ first-neighbor planes have been superposed such that
the $z$ axes remain parallel and the 
$\bC_m$ coincide in a single point $\bC$,
taken here as the origin of the coordinate system.
The dots represent points of polar coordinates
$(r_m,\theta_m)$, defined in figure \ref{fig_SSF}.
In order to symmetrize the figure
the points $(r_m,-\theta_m)$ are also shown.
The hull of a spindle torus,
indicated by the circular arc, becomes clearly visible. 
See text. 
}
\label{fig_rhotheta}
\end{figure}

Figure \ref{fig_rhotheta} is based on the same set of five
$18$-edged cell faces. With  each face there are associated $18$
planes of the type shown in figure \ref{fig_SSF}, 
each one passing through the two focal seeds and through 
one first neighbor seed $\bF_m$. 
In figure \ref{fig_rhotheta} we have superposed these 
$5\times 18$ planes such that the points $\bC_m$
coincide in a single point called $\bC$
(this blurs of course the positions of the focal seeds). 
The positions $(r_m,\theta_m)$ with respect to $\bC$,
defined in figure \ref{fig_SSF}, 
of the first neighbor seeds $\bF_m$  are shown.
The figure clearly shows the appearance of the hull of a spindle torus,
indicated by the circular arc. We have chosen in this figure a radius
$\sfrh_{\rm av}$ as well as
somewhat arbitrary values for $L_{\rm av}$ and $\sfRh_{\rm av}$
such as to obtain a good visual fit.
We now recall the discussion of section \ref{sec_sphericity}
that concerned the spherical surface:
here, in a fully analogous way, the scatter of the dots about the arc is
a measure of the combined effect of
the shell width $w_n$, determined in section \ref{sec_facetheory},
and the elastic deformations, left unstudied, of the toroidal surface. 
The scarcity of points as one approaches the axis of revolution is 
an effect of diminishing phase space.

\subsection{Average area $A_n$ and perimeter $P_n$}
\label{sec_AnPn}

\begin{figure}
\begin{center}
\scalebox{.40}
{\includegraphics{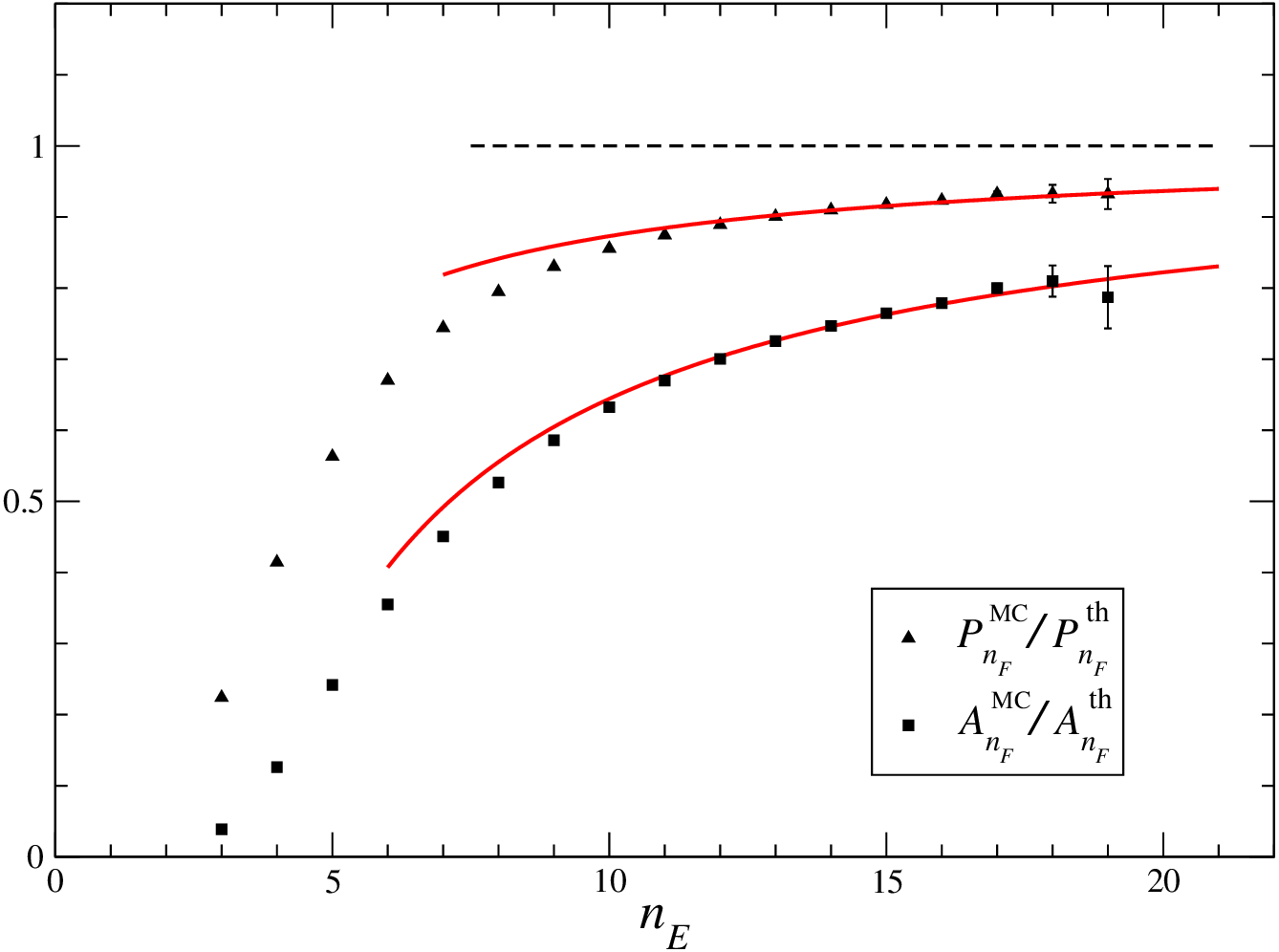}}
\end{center}
\caption{\small Monte Carlo 
averages $A_\nE^{\rm MC}$ and $P_\nE^{\rm MC}$ of the area and
perimeter, respectively,
of an $\nE$-edged cell face, each divided by its theoretical asymptotic
behavior, Eqs.\,(\ref{resAnPn}). 
Both sets of data points are predicted, therefore, to tend to unity as
$\nE\to\infty$. The solid red lines approach this limit value
as  $\sim n^{-1}$ and represent our best estimates for the
next-order correction to the leading
asymptotic behavior (section \ref{sec_higher}).}
\label{fig_face}
\end{figure}

In figure \ref{fig_face}
we have represented our Monte Carlo averages $A_\nE^{\rm MC}$
and $P_\nE^{\rm MC}$ for the area and perimeter, respectively,
of the $\nE$-edged cell face,
averaged over the set of $4\times 10^9$ cells.
Each quantity has been divided by its theoretical large-$\nE$
behavior (\ref{resAnPn}), so that for both 
the data points are expected to tend
to unity as $\nE\to\infty$. 
We emphasize again that the theory has no adjustable parameters. 
The data for $A_\nE^{\rm MC}$ and $P_\nE^{\rm MC}$ 
appear to fully conform to the theoretical prediction,
even if the finite-$\nE$ corrections are still large. 
We will analyze these subleading terms to the asymptotic laws
in section \ref{sec_higher}.

\subsection{Focal distance $L$}
\label{sec_2L}

As far as we are aware, the statistics of the focal distance $L$
for given edgedness $\nE$ has not hitherto received any attention 
in the literature, whether it be its average $L_\nE$ or its full 
probability distribution  $Q_\nE(L/L_\nE^{\rm th})$. 
The theoretical result of Eq.\,(\ref{resLn}) for $L_\nE$ 
is not intuitive and 
it is therefore of utmost importance that we compare the predictions
(\ref{resLn}) and (\ref{xQy}) to the Monte Carlo data.

\begin{figure}
\begin{center}
\scalebox{.45}
{\includegraphics{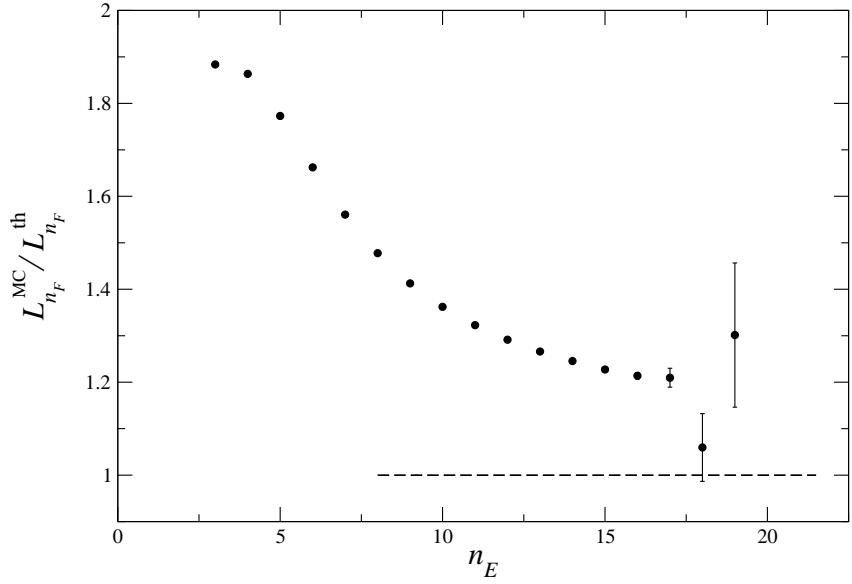}}
\end{center}
\caption{\small 
Monte Carlo average $L_\nE^{\rm MC}$ of the
focal distance  divided by its 
theoretical asymptotic behavior (\ref{resLn}).
The data points are, therefore, predicted to tend to unity as $\nE\to\infty$.}
\label{fig_Ln}
\end{figure}

In figure \ref{fig_Ln} we have represented the Monte Carlo average
$L_\nE^{\rm MC}$, divided by its  theoretical large-$\nE$ behavior 
(\ref{resLn}),
so that the data points are expected to tend to unity for
$\nE\to\infty$. 
The Monte Carlo data 
are fully compatible with the asymptotic limit value,
even though there appear, here as before, sizeable finite-$\nE$ corrections. 

\begin{figure}
\begin{center}
\scalebox{.45}
{\includegraphics{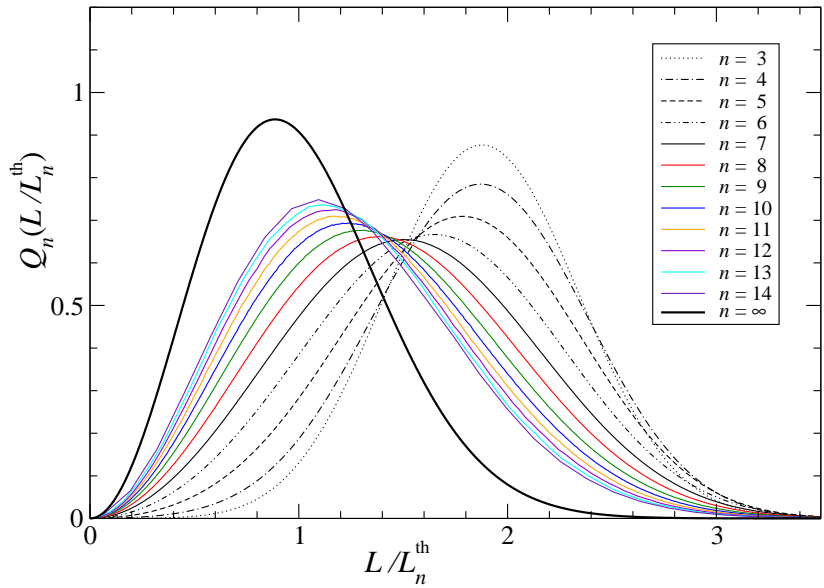}}
\end{center}
\caption{\small 
Monte Carlo data for the probability distributions $Q_n(L/L_n^{\rm th})$ 
of the focal distance $L$. The heavy black curve is the theoretical limit
distribution $Q(y)$ of Eq.\,(\ref{xQy}).}
\label{fig_QnL}
\end{figure}

\begin{figure}
\begin{center}
\scalebox{.45}
{\includegraphics{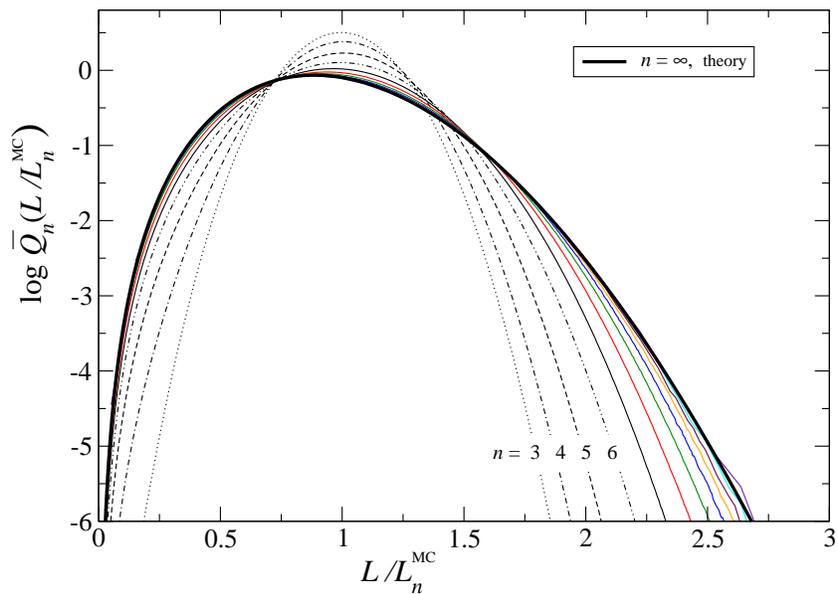}}
\end{center}
\caption{\small
Monte Carlo data for the
logarithm of the scaled probability distributions $\bar{Q}_n(L/L_n^{\rm MC})$
of the focal distance $L$. The color code is as in figure \ref{fig_QnL};
the curves for $n=3,4,5,6$ have been labeled explicitly.
These distributions all have unit average.
The heavy black curve is the theoretical limit distribution
$\log Q(y)$ of Eq.\,(\ref{xQy}).}
\label{fig_logQnL}
\end{figure}

In figure \ref{fig_QnL} we proceed to a more detailed comparison. This
figure shows, for $n=7$ through $n=14$,
the distributions $Q_n(L/L_n^{\rm th})$ of the scaled variables
$L/L_n^{\rm th}$.
We constructed this figure
by collecting the values of $L$
for each $n$ separately in bins of width $0.005$.
In order to suppress fluctuations,
we combined for the larger $n$ values
groups of neighboring bins into larger
ones: for $n=11,12,13,14$
we grouped together  $2,4,8,16$ of the original bins, respectively.
There is a clear tendency for the $Q_n(y)$ to approach the theoretical
limit distribution.

In figure \ref{fig_logQnL}
we investigate the {\it shape\,} of the distributions $Q_n(y)$.
Let $\alpha = L_n^{\rm th}/L_n^{\rm MC}$ and define
rescaled distributions $\bar{Q}_n(L/L_n^{\rm MC})= \alpha Q_n(y)$,
which have unit average. 
We have plotted the $\bar{Q}_n$ semilogarithmically to allow for
comparisons over a wider range of the abscissa. 
It appears that the shape of the $\bar{Q}_n$ converges 
rapidly to the
theoretically predicted limit given by Eq.\,(\ref{xQy}).
Hence the limiting shape of the distribution is attained well
before the average reaches its limit value.
This excellent agreement comes somewhat as a
surprise since we had no specific reasons beforehand to expect it. 

In any case, the Monte Carlo data for $L$
provide ample evidence of the fact that
$\LnE/\sfrh_\nE \to 0$  as   $\nE\to\infty$,
and that therefore the limit torus 
has equal major and minor radii: it is 
a true doughnut but with a hole of zero diameter.


\section{Higher order terms}
\label{sec_higher}

\begin{figure}
\begin{center}
\scalebox{.40}
{\includegraphics{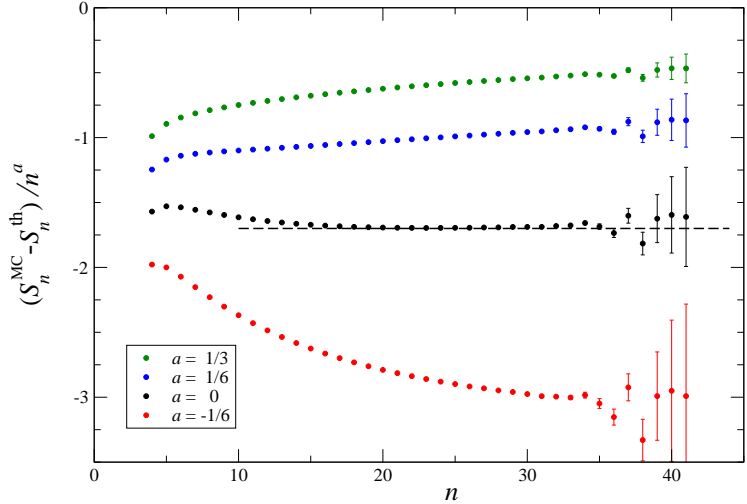}}
\end{center}
\caption{\small Trying to fit the next-to-leading term in the
  asymptotic expansion of $S_n$ by different powers $a$. From top to
  bottom $a=\tfrac{1}{3}, \tfrac{1}{6}, 0, -\tfrac{1}{6}$.}
\label{fig_surnext}
\end{figure}

\begin{figure}
\begin{center}
\scalebox{.40}
{\includegraphics{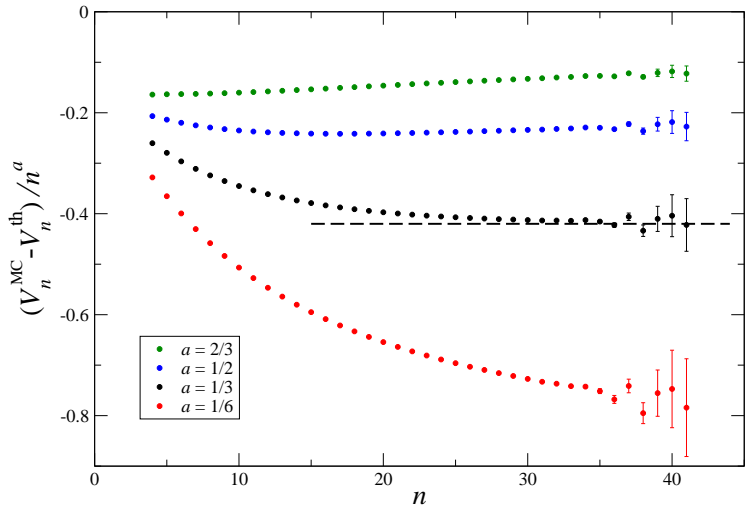}}
\end{center}
\caption{\small Trying to fit the next-to-leading term in the
  asymptotic expansion of $V_n$ by different powers $a$. From top to
  bottom $a=\tfrac{2}{3}, \tfrac{1}{2}, \tfrac{1}{3}, \tfrac{1}{6}$.}
\label{fig_volnext}
\end{figure}

We will let $n$ stand for either $\nE$ or $\nF$, and
$X_n$ for any of the four quantities $V_n, S_n, A_n$, and
$P_n$ studied in the preceding sections.
We there determined their leading large-$n$ behavior
$X_n^{\rm th} \simeq c_0 n^{a_0}$,
and now ask if we can go beyond that.
Each of these averages
presumably has an asymptotic expansion in powers $n$ of the form
\beq
X_n = c_0n^{a_0} + c_1n^{a_1} + \ldots, \qquad n\to\infty,
\label{asptexpX}
\eeq
with coefficients $c_1,c_2,\ldots$ and powers $a_1,a_2,\ldots$
of which we have no theoretical knowledge.
We will nevertheless rely
on the idea that the only powers that one may reasonably expect 
are powers of $n^{1/3}$.
We will try to determine these from the Monte Carlo data.
Our procedure will follow the definition of an asymptotic expansion:
We plot $(X_n^{\rm MC}-X_n^{\rm th})/n^a$ 
for selected values of $a$ and 
look for the $a$ that makes this quantity tend to a constant
when $n$ gets large. That value of $a$ is then equal to $a_1$ and the
constant is equal to $c_1$.
How well this works depends in part on the accuracy of
the simulation data,
and in part on whether we are sufficiently far in the asymptotic regime,
a question to which we have no certain answer.

Let us consider first the $n$-faced cell.
The most clearcut case is provided by its surface area $S_n$, plotted in
figure \ref{fig_surnext} 
for a selection of values of $a$ that also include half-integer powers of
$n^{1/3}$. This plot
seems to clearly single out $a=a_1=0$ as the next
exponent in the series (\ref{asptexpX}) for $X_n=S_n$.
Accepting this exponent value we are led to conclude that the
corresponding constant takes the value $c_1=-1.70$,
indicated by the horizontal dashed line in the figure.
In figure \ref{fig_volnext}
a similar analysis has been performed for $V_n$. 
It points towards an exponent $a_1=1/3$ and a
coefficient $c_1=-0.42$.
The resulting two-term asymptotic series for $V_\nF$ and $S_\nF$
have been listed in table \ref{table2}.
The curve representing the subleading term 
has been drawn in figure \ref{fig_cell} for both quantities.

\begin{table}
\begin{center}
\begin{scriptsize}
\begin{tabular}{|lllc|}
\hline
Quantity & Symbol  & Leading term(s) for large $n$ & Note\\
\hline
Average surface area of an $n$-faced 3D cell  & $S_n$ &
$({9\pi}/{16})^{1/3}n^{2/3} - 1.70$ & {\it a}\\
Average volume of an $n$-faced 3D cell        & $V_n$ &
$n/8 - 0.42 n^{1/3}$ & {\it a}\\
Average perimeter of an $n$-edged face of a 3D cell & $P_n$&   
$(4\pi)^{1/3}n^{1/3} - 2.95 n^{-2/3}$  & {\it a}\\
Average area of an $n$-edged face of a 3D cell & $A_n$ &
$(4\pi)^{-1/3}n^{2/3} - 1.53 n^{-1/3}$ & {\it a}\\    
Average of the distance $L$ between the seeds of  && &\\
${}$\hspace{5mm} two 3D cells sharing an $n$-edged face  & $L_n$ &
$2^{5/3}3^{-1/2}\pi^{-7/6}n^{-1/6}$    & {\it b}\\
Probability distribution of $y=L/L_n$              & $Q(y)$  &
$32\pi^{-2}y^2\exp(-4y^2/\pi)$ & {\it b}\\
Average perimeter of an $n$-sided 2D cell & $P^{(2)}_n$ &
$\pi^{1/2}n^{1/2} - (5/8)\pi^{1/2}n^{-1/2}$ & {\it c}\\
Average area of an $n$-sided 2D cell          & $A^{(2)}_n $ &
$n/4 - 0.6815$ & {\it c}\\
\hline
\end{tabular}
\end{scriptsize}
\end{center}
\vspace{-3.5mm}

\begin{scriptsize}
\noindent $^a$ This work. 
First term from large-$n$ theory, expected to be exact; 
second term fitted.\\[-2mm]
\noindent $^b$ This work. Leading order term from large-$n$ theory.\\[-2mm]
\noindent $^c$ First term analytically exact \cite{Hilhorst05b}; 
second term from a high precision fit \cite{Hilhorst07}.
\end{scriptsize}
\caption{\footnotesize
Summary of predictions for the asymptotic large-$n$ behavior
of several quantities associated with Poisson-Voronoi tessellations. 
The last two lines concern earlier work.}
\label{table2}
\end{table}

\begin{figure}
\begin{center}
\scalebox{.40}
{\includegraphics{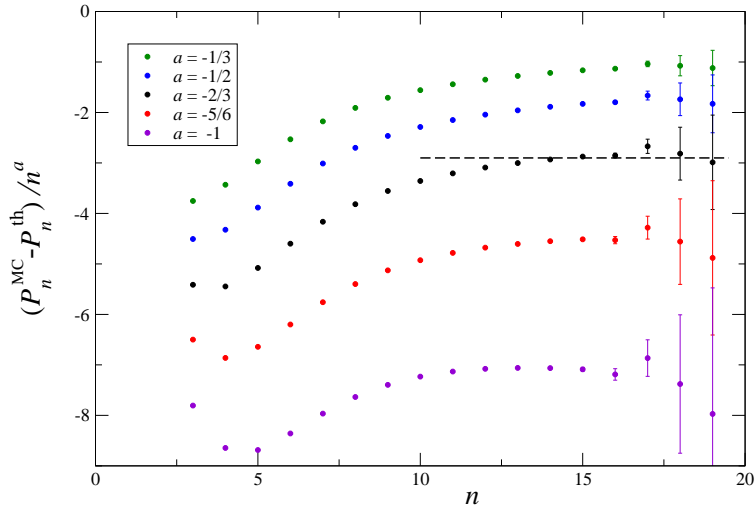}}
\end{center}
\caption{\small Trying to fit the next-to-leading term in the
  asymptotic expansion of $P_n$ by different powers $a$. From top to
  bottom $a=-\tfrac{1}{3}, -\tfrac{1}{2}, -\tfrac{2}{3},
  -\tfrac{5}{6}$, -$1$.}
\label{fig_pernext}
\end{figure}

\begin{figure}
\begin{center}
\scalebox{.40}
{\includegraphics{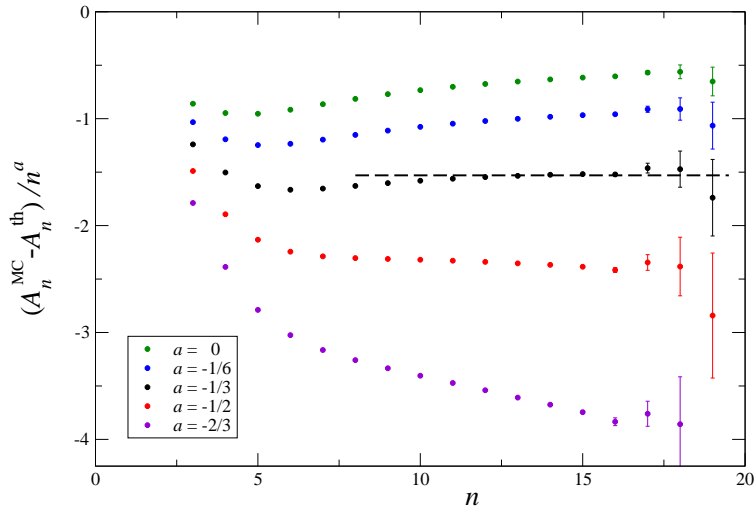}}
\end{center}
\caption{\small Trying to fit the next-to-leading term in the
  asymptotic expansion of $A_n$ by different powers $a$. From top to
  bottom $a=0,-\tfrac{1}{6}, -\tfrac{1}{3}, -\tfrac{1}{2},
  -\tfrac{2}{3}$.}
\label{fig_arenext}
\end{figure}

Let us next consider the average perimeter $P_n$ and area $A_n$ of an
$n$-edged face. Figures \ref{fig_pernext} and
\ref{fig_arenext} show the attempts to fit the asymptotic behavior.
The evidence is less convincing here than
for the case of the cell volume and surface area, and 
it certainly helps to assume at this point
that the exponents are quantized as
multiples of $1/3$. The values $a_1=-2/3$ for $P_n$ and $a_1=-1/3$ for
$A_n$ appear to best fit the data, and accepting these 
we obtain estimates for the coefficients, again indicatd by horizontal
dashed lines.
The resulting two-term asymptotic series for $P_n$ and $A_n$
have also been listed in table \ref{table2}. 
The curve representing the subleading term 
has been drawn in figure \ref{fig_face} for both quantities.

\section{Discussion}
\label{sec_discussion}

We have summarized
the main results of this paper in table \ref{table2}. 
For comparison the two bottom lines
in this table show analogous results
obtained earlier \cite{Hilhorst05b,Hilhorst07} 
for the average perimeter $P^{(2)}_n$
and area $A^{(2)}_n$ of a two-dimensional Poisson-Voronoi cell.
The status of these results, briefly indicated in the
notes at the bottom of the table, is as follows.
We basically have two reasons to believe that in three dimensions
the results from large-$n$ theory are exact for 
the four quantities $V_n$, $S_n$, $A_n$, and $P_n$. 
The first reason is that in two dimensions this theory 
reproduces the exactly known leading order
results for $A_n^{(2)}$ and $P_n^{(2)}$. The second one is that the
theory leads to what looks like a sound basic principle:
The probability of occurrence  (entropy) of an ``event'' 
imposing restrictions on the positions of $n$ seeds
is maximized by displacing
(with respect to a random configuration) 
only those $n$ seeds,
thus evacuating a spatial region of volume $n/\lambda$ (where $\lambda$
is the seed density).
For the $n$-faced cell
this region is a sphere [Eq.\,(\ref{relRncell})], 
for the $n$-edged face it is a torus [Eq.\,(\ref{solrhox1})]
with major and minor radii that for $n\to\infty$ become equal.

Large-$n$ theory, at least in its present form, does not allow for a
systematic expansion of the averages considered above in negative powers
of $n$. We have therefore based our determination of the correction
terms on fits of the Monte Carlo data,
guided by theoretical considerations.
In next-to-leading order
there is in each case a power of $n$ and a coefficient to estimate. 
In the case of $V_n$ and $S_n$ these come out fairly
unambiguously.
In the case of $A_n$ and $P_n$ we have been led, in addition, by a
certain systematics that appears: just like  $A_n^{(2)}$ and $P_n^{(2)}$
in two dimensions, 
and for reasons that we do not at this point fully understand,
the correction terms for $A_n$ and $P_n$  turn out to differ from the 
leading order behavior by integer powers of $n^{-1}$.

The focal distance $L$ is a quantity that enters in a different way
into the theory. First, in contradistinction
to the four averages discussed above,
its theoretical mean value $L^{\rm th}_n$ 
does not diverge with growing $n$  but tends to zero as $\sim n^{-1/6}$.
The Monte Carlo data for $L^{\rm MC}_n$ 
are fully compatible with this prediction;
there are again substantial finite-$n$ corrections which,
in this quantity, we  have not attempted to estimate.
Secondly, it appears that even for large $n$ 
the probability distribution $Q_n$ of the scaled variable
$y=L/L^{\rm th}_n$ does not become sharply peaked but 
approaches a well-defined limit law $Q(y)$ [Eq.\,(\ref{xQy})].
Although we had no {\it a priori\,} indication 
about the reliability of these conclusions from large-$n$ theory,
the distribution $Q(y)$ appears to be in excellent agreement with theory.

From the theoretical point of view it is worthwhile to recall an invariance
property exploited in section \ref{sec_wn}, 
{\it viz.} the fact that a cell face does not change when
any or all of the first neighbors (to its two focal seeds)
are rotated over arbitrary angles in their `first-neighbor' planes.
We suspect that this invariance may open the road to an exact
determination of the properties of the many-sided cell face. 


\section{Conclusion}
\label{sec_conclusion}

We have performed and theoretically analyzed
Monte Carlo simulations of three-dimensional
Poisson-Voronoi cells. The number of cells generated, namely
equals $4\times 10^9$, is larger than in all earlier work.
Our method of analysis has been the heuristic `large-$n$' theory, 
applicable to Voronoi cells with a large
number $\nF$ of faces, and to cell faces with a large number $\nE$ of
edges. The latter application has required a substantial extension of
the theory that we describe in this paper.
Whereas many-faced cells must be analyzed in terms of a spherical
geometry, we found that the many-edged cell face
requires the geometry of a spindle torus.
The squared major and minor radii of that torus differ by $L^2$,
where the `focal' distance $L$ is half the distance 
between the seeds of the two cells sharing that face.
We were natuarally led to investigate
the statistics of $L$ and found again good agreement between theory
and Monte Carlo data.

The results presented here highlight, in addition, the potential use of Monte
Carlo simulations in conjunction with large-$n$ theory as a means of gaining
insight into the properties of 3D Poisson-Voronoi cells. 




\appendix

\end{document}